\documentclass[11pt,english]{article}
\usepackage{rotating}
\usepackage{axodraw}
\usepackage{latexsym}
\usepackage{graphicx}
\usepackage{psfrag}
\usepackage{amsmath,amssymb}
\makeatletter
\def\section{\@startsection {section}{1}{\z@}{-3.5ex plus -1ex minus
 -.2ex}{2.3ex plus .2ex}{\large\bf}}
\def\subsection{\@startsection{subsection}{2}{\z@}{-3.25ex plus -1ex
minus -.2ex}{1.5ex plus .2ex}{\normalsize\bf}}
\makeatother
\makeatletter

\@addtoreset{equation}{section}

\makeatother

\newcommand{\captionfonts}{\small}
\makeatletter  
\long\def\@makecaption#1#2{%
  \vskip\abovecaptionskip
  \sbox\@tempboxa{{\captionfonts #1: #2}}%
  \ifdim \wd\@tempboxa >\hsize
    {\captionfonts #1: #2\par}
  \else
    \hbox to\hsize{\hfil\box\@tempboxa\hfil}%
  \fi
  \vskip\belowcaptionskip}
\makeatother   
\def\pa{\partial}
\def\dslash{\raisebox{1pt}{$\slash$} \hspace{-6pt} \partial}

\def\bea{\begin{eqnarray}} \def\eea{\end{eqnarray}}
\def\be{\begin{equation}} \def\ee{\end{equation}} \def\nn{\nonumber}
\def\a{& \hspace{-7pt}} \def\c{\hspace{-5pt}} \def\Z{{\bf Z}}
 \def\ov{\overline}

\def\ov{\over}
\def\ha{{1\over 2}}

\newcommand{\promille}{%
  \relax\ifmmode\promillezeichen
        \else\leavevmode\(\mathsurround=0pt\promillezeichen\)\fi}
\newcommand{\promillezeichen}{%
  \kern-.05em%
  \raise.5ex\hbox{\the\scriptfont0 0}%
  \kern-.15em/\kern-.15em%
  \lower.25ex\hbox{\the\scriptfont0 00}}

\begin{document}

\thispagestyle{empty}

\begin{center}
\hfill SISSA-32/2009/EP \\

\begin{center}

\vspace{1.7cm}

{\LARGE\bf  Renormalization group in Lifshitz-type theories}

\end{center}

\vspace{1.4cm}

{\bf  Roberto Iengo$^1$, Jorge G. Russo$^2$ and Marco Serone$^1$}

\vspace{1.2cm}

${}^1${\em ISAS-SISSA and INFN, Via Beirut 2-4, I-34151 Trieste, Italy} \\

\medskip

${}^2${\em Instituci\'o Catalana de Recerca i Estudis
Avan\c cats (ICREA),\\
Departament ECM and Institut de Ciencies del Cosmos, \\
Facultat de F\'{\i}sica - Universitat de Barcelona,\\
Diagonal 647, E-08028 Barcelona, Spain.\\ }

\end{center}

\vspace{0.8cm}

\centerline{\bf Abstract}

We study the one-loop renormalization and evolution of the couplings in 
scalar field theories of the Lifshitz type, i.e. with different scaling in space and time.
These theories are unitary and renormalizable, thanks to higher spatial derivative terms that modify the particle propagator at high energies, but at the expense of explicitly breaking Lorentz symmetry. 
We study if and under what conditions the Lorentz symmetry can be considered as emergent at low energies by studying the RG evolution
of the  ``speed of light'' coupling $c^2_\phi$ and, for more than one field,  
of $\delta c^2\equiv c^2_{\phi_1}-c^2_{\phi_2}$ in simple models. We find that in the UV both $c^2_\phi$ and $\delta c^2$ generally flow logarithmically with the energy scale. A logarithmic running of $c^2$ persists also at low-energies, if $\delta c^2 \neq 0$ in the UV.
As a result, Lorentz symmetry is not recovered at low energies
with the accuracy needed to withstand basic experimental constraints, 
unless all the Lorentz breaking terms,  including $\delta c^2$,  
are unnaturally fine-tuned to extremely small  values in the UV. We expect that the considerations of this paper
will apply to any generic  theory of Lifshitz type, including a recently proposed
quantum theory of gravity by Ho$\breve{{\rm r}}$ava.

\vspace{2 mm}
\begin{quote}\small

\end{quote}

\vfill

\newpage

\section{Introduction}

 Recently, there has been an increasing interest in non-relativistic quantum field theories where Lorentz invariance is explicitly broken at high energies and hopefully recovered at low energies. In particular, in \cite{Anselmi:2007ri,Anselmi2,Anselmi:2008bt} (see also \cite{Dhar:2009dx}),  general gauge theories, including  non-relativistic extensions of the Standard Model,  were proposed and investigated, while in \cite{Horava1,Horava2} similar constructions were implemented in  Yang-Mills theories in 4+1 space-time dimensions and  membrane theory. The same type of construction was then extended to four-dimensional quantum gravity in \cite{Horava3},\footnote{See also \cite{Lu:2009em,Charmousis:2009tc} for a (partial) list of further works that investigate the proposal of \cite{Horava3}.}  where it was suggested  that the resulting theory may provide a candidate for a renormalizable and unitary quantum theory of gravity which flows in the infrared (IR) to Einstein theory.
 
The ultraviolet (UV) behavior of all these theories is substantially ameliorated by the presence of higher derivative  (in the spatial directions only) quadratic terms that improve the UV behavior of the particle propagator, without introducing ghost-like degrees of freedom that 
in Lorentz invariant higher derivative theories typically spoil the unitarity of the theory.
The proposed theories are of Lifshitz type \cite{Lif}. In the UV, they exhibit, at the classical level, an anisotropic scaling symmetry under which time and space scale differently: $t = \lambda^z t^\prime$,
$ \vec{x} = \lambda \vec{x}^{\prime}$, where $z$ is the critical exponent, equal to one in a Lorentz invariant theory. The renormalizability properties of these theories have been extensively studied in \cite{Anselmi:2007ri,Anselmi2} for scalar, fermion and gauge theories.  
The usual power-counting argument for the renormalizability of a theory does not strictly hold anymore, but it is essentially still valid, provided one substitutes the standard scaling dimensions of the operators  by their ``weighted scaling dimensions'' \cite{Anselmi:2007ri}, i.e. by the
dimensions implied by the assignment $[x]_w=-1$ and $[t]_w=-z$. 
 Lifshitz-type theories exhibit at least two qualitative different energy regimes, set by the scale (denoted by $\Lambda$)
of the higher derivative operators. We will generally denote as UV regime the energy  range  $E\gg \Lambda$, where
the theory is manifestly non-Lorentz invariant. This is the proper ``Lifshitz" regime, where the effective scaling dimensions of the operators are the weighted ones. We denote as IR the range $E\ll \Lambda$, where
the theory is expected to smoothly reach the ``standard''  regime,  where the operators are classified by their standard scaling dimensions. Weighted relevant operators break  the anisotropic scaling symmetry explicitly and, at low energies, the theory is expected to flow to the Lorentz invariant theory with $z=1$. 
This is, however, a  non-trivial (and obviously crucial) point, since there is no dynamical principle for which Lorentz symmetry should emerge in the IR.
As a matter of fact, we find that Lorentz invariance is recovered in the IR only if an unnatural fine tuning 
of the parameters of the theory ensure that all sources of Lorentz violation in the theory are tiny and below the rather stringent, presently known experimental bounds.
These issues, expected on general grounds \cite{Collins:2004bp}, are explored by performing a one-loop calculation in concrete, simple, models, 
which also clarify the IR--UV structure of Lifshitz-type theories.

More specifically, the purpose of this paper is to study the Renormalization Group (RG) evolution at one-loop level of  simple scalar field theories of Lifshitz type. In particular, we will calculate
the one-loop beta-functions of the weighted marginal operators in the theory and solve the corresponding equations to study the   evolution of the associated couplings $g_i$.
Once this is performed, we will focus our attention on a particular weighted relevant operator,
$c_\phi^2 (\vec{\partial} \phi)^2$, and study the RG evolution of the ``speed of light" parameter $c_\phi$ for $\phi$
(physically, in the low energy theory $c_\phi $ represents the maximal speed for $\phi $ particles).
In order to keep the technical analysis as simple as possible, we will mostly consider scalar field theories in higher dimensions with $z=2$, namely a $\phi^6$ theory in $D=4$ spatial dimensions, with quartic derivative couplings as well, and a $\phi^3$ theory in $D=10$ spatial dimensions. These two theories are among the simplest
theories which are i) of Lifshitz type, ii) their $\beta$-functions are respectively positive and negative in the UV,  iii) $c_\phi^2$ has a non-trivial running already at one-loop level. These theories are obviously toy laboratories (in particular, the $\phi^3$ theory is not even stable), yet they manifest, in a simple context, the main features that more complicated and ``realistic" models of this sort should exhibit. In both theories, $c_\phi^2$ typically 
shows a logarithmic running in the range $E\gg \Lambda$, where $\Lambda$ is the high-energy scale where the theory is anisotropic:
 \be
c^2_\phi(E) = c^2_\phi(E_0) \bigg[ 1 + f  \log\Big(\frac{E}{E_0}\Big) \bigg]^{n_\phi}\,,
\label{cIntro}
\ee
here 
$n_\phi$ an ${\cal O}(1)$ particle-dependent constant and $E_0\gg \Lambda$ a reference scale in the UV range. In eq.(\ref{cIntro}),
we schematically denote by $f$  the radiative coefficient governing the RG flow, which depends
on the coupling constants of the weighted marginal operators.
It is also important to investigate what happens in the presence of more than one field, 
and particularly if and under what conditions their ``speed of light" parameters converge to the same value.
To address this question, we have also studied the RG evolution of the difference $\delta c^2 = c_{\phi_1}^2-c_{\phi_2}^2$. Under the assumption that $\delta c^2\ll1 $, one schematically finds, for $E\gg \Lambda$, 
\be
\delta c^2(E) = \delta c^2(E_0) \bigg[ 1 +  f^\prime  \log\Big(\frac{E}{E_0}\Big) \bigg]^{n_\delta} + \delta g(E_0)\bigg\{\bigg[ 1 +  f^{\prime\prime} \log\Big(\frac{E}{E_0}\Big)\bigg]^{n_g} -1 \bigg\} \,,
\label{deltacIntro}
\ee
where $\delta g$ are small perturbations around some fixed-point solutions of the RG evolution.
Eqs.(\ref{cIntro}) and (\ref{deltacIntro}) summarize the quantum evolution of $c^2$ and $\delta c^2$ in the UV regime, as given by the
marginal couplings.

After having studied the UV, we move on  to analyze the IR regime $E\ll \Lambda$. 
We will see that $\Lambda$ is the characteristic scale below which Lifshitz theories  turn into ``standard theories". More precisely, we will explicitly show that  the RG evolution of all  weighted marginal couplings is essentially frozen below $\Lambda$, in complete analogy to the decoupling of a massive particle in a standard quantum field theory. The key point is, of course, that in the IR the relevant propagator term is the usual one, quadratic in the momentum, while the higher derivative terms can be neglected (it is however a delicate point, given that the theory with the usual quadratic propagator is non-renormalizable).
Taking into account only the effect of  weighted marginal couplings,  for $E\ll \Lambda$ we find
 \be
c^2_\phi(E) = c^2_\phi(\Lambda) \bigg[ 1 + {\cal O} \Big(\frac{E^{2}}{\Lambda^{2}} \Big) \bigg]\,, \ \ \ \ 
\delta c^2(E) = \delta c^2(\Lambda) \bigg[ 1 + {\cal O} \Big(\frac{E^{2}}{\Lambda^{2}} \Big) \bigg]\,,
\label{cIntro2}
\ee
which shows that, for sufficiently high $\Lambda$, the IR effect of the weighted marginal couplings can be neglected.
This is expected, since in the IR the usual classification of operators in terms of canonical rather than weighted dimensions holds. What is marginal in the UV becomes then irrelevant in the IR. We will explicitly show how the $\beta$-functions smoothly change their behavior
going from the UV to the IR by computing them in a momentum subtraction renormalization scheme, where all the decoupling effects are manifest. 

However, care has now to be paid for the weighted relevant operators which become standard marginal in the IR, since they can efficiently mediate  the UV Lorentz violation to the low-energy theory. Indeed, we will show, by explicitly working 
out a toy example in 3+1 space-time dimensions, that a logarithmic running like eq.(\ref{cIntro}) (with $E_0$ replaced now by $\Lambda$) still holds in the IR, with $f$ depending now on the (standard) relevant couplings and being proportional to any Lorentz symmetry breaking coefficient of the low-energy effective theory, remnant of the
Lifshitz-like nature of the UV completion.

In general, then, Lorentz symmetry is {\it not} emergent in the IR in theories of Lifshitz type. Recovering Lorentz symmetry 
 would require some dynamical principle keeping all sources of Lorentz violation sufficiently small. 
The experimental bounds on $\delta c^2$ for ordinary particles are of the order of $10^{-(21\div 23)}$  \cite{Coleman:1998ti}, which give an idea of the order of magnitude of the fine--tuning that is needed. An indirect bound on $\delta c^2$ for any charged particle is implicitly given by eq.(\ref{cIntro}).
In the case of photons, for instance, an experimental constraint on the energy dependence of $c_\gamma^2$  by the FERMI experiment \cite{Abdo:2009zz} gives,  taking $n_\gamma \sim 1$ in eq.(\ref{cIntro}),  the following rough constraint on $f$:
\be
|f| \lesssim 10^{-16}\,. 
\label{cgammaConstr}
\ee
Modulo a loop-factor and coefficients of order one, the bound (\ref{cgammaConstr}) can be seen as a bound on $\delta c^2$ for any charged particle.
We expect that these fine-tuning problems will affect all generic quantum field theories of Lifshitz type, in particular,
the non-relativistic standard model of \cite{Anselmi:2008bt} and the proposed quantum gravity theory of \cite{Horava3}. In the latter case,
after coupling the theory to matter, the  problems mentioned above will reappear for the Standard Model particles, where parameters like $\delta c^2$ have tight experimental constraints. 

The plan of this paper is as follows. In section 2 we briefly review the main properties of the Lifshitz-like
theories, using, for the sake of illustration, a scalar field theory in 3+1 dimensions. In section 3.1 we study the one-loop renormalization of a single scalar field 
$\phi^6$ theory with derivative $\phi^4$ interactions in $D=4$ spatial dimensions; in section 3.2 this analysis is extended to the case
of two coupled fields. In section 4.1 we study the one-loop renormalization of a single scalar field $\phi^3$ in $D=10$; the two-field case is dealt with in section 4.2. The analysis in sections 3 and 4 are performed  using
the minimal subtraction (MS) scheme, suitable for finding the $\beta$-functions in the UV regime ($E\gg \Lambda$).
 In section 5 we study the IR behavior ($E\ll \Lambda$) of Lifshitz-like theories by using the momentum subtraction scheme. After reviewing in section 5.1 the IR behavior  of the
$\beta$ function in conventional
 $\phi^4$- theory,  
in section 5.2 we show that the $\beta$--functions produced by weighted marginal couplings go to zero for $E\ll \Lambda$.   
In section 5.3 we discuss the contribution of the weighted relevant operators (marginal in the standard sense) on  Lorentz symmetry breaking effects in the IR. 
We will see that the presence of a non-zero $\delta c$, inherited from the UV, induces a running in $c^2$ at low-energies. The effect is general
and we expect that it should apply to any low-energy field theory (i.e. not only to theories of Lifshitz type) perturbed by Lorentz symmetry breaking terms.
In section 6 we conclude.

\section{Renormalizable Lifshitz-like scalar field theories} 

Unconventional scalar field theories of the Lifshitz type, with higher derivative 
interactions and higher derivative quadratic terms, have been extensively studied in \cite{Anselmi:2007ri}. Here we 
briefly review some aspects of the construction that will be useful in what follows and refer the reader to \cite{Anselmi:2007ri} for more details. As mentioned in the introduction, the key point of the whole construction is to break Lorentz invariance, so that one is allowed to introduce higher derivative
terms in the spatial derivatives and quadratic in the fields, without necessarily introducing
the dangerous higher time derivative terms  that would lead to violations  of unitarity. 
In doing so, the UV behavior of the propagator is improved and theories otherwise non-renormalizable
become effectively renormalizable. A useful guiding principle to easily classify and identify
the renormalizable theories in this enlarged set-up is achieved by demanding an invariance under ``anisotropic"\footnote{The word ``anisotropic" arises from condensed-matter physics.
In all the instances we consider, we assume to be in a so-called  ``preferred frame'' \cite{Coleman:1998ti} where spatial $SO(3)$ rotations, translations and time-reversal are unbroken symmetries.} scale transformations:
\be
t = \lambda^z t^\prime \,, \ \ \ \ \ x^i = \lambda x^{i\prime}\,, \ \ \ \ \
\phi(x^i,t) = \lambda^{\frac{z-D}{2}} \phi^\prime(x^{i\prime},t^\prime)\,,
\label{scalingW}
\ee
where $i=1,\ldots, D$ parametrizes the spatial directions. The parameter $z$ is known as ``critical exponent" and, when it equals one, the transformations (\ref{scalingW}) reduce to the usual Lorentz invariant scale transformations. According to eq.(\ref{scalingW}), we can assign to the coordinates and to the fields a ``weighted'' scaling dimension as follows:
\be
[t]_w = - z\,, \ \ \ \  [x^i]_w = -1\,, \ \ \ \  [\phi]_w = \frac{D-z}{2}\,.
\ee
It is straightforward to see that at the quadratic level, modulo  total derivative terms, the Lagrangian for a single scalar field, invariant under (\ref{scalingW}), reads
\be
{\cal L}_{quad} =\frac 12\dot\phi^2-\frac{a^2}{2\Lambda^{2(z-1)}}(\partial_i^z \phi)^2\,,
\label{Lquad}
\ee
with  $a^2$ being a dimensionless coupling and $\Lambda$  a high-energy  scale parametrizing the strength of the higher derivative operator.  Due to the improved UV behavior of the propagator resulting from (\ref{Lquad}) when $z>1$, the usual power-counting argument for the renormalizability of a theory is no longer applicable. The required modification is obtained by substituting the scaling dimensions of the operators  by their ``weighted scaling dimensions'' \cite{Anselmi:2007ri}.\footnote{Sometimes in the literature the weighted scaling dimension is introduced as the standard scaling dimension in some non-standard natural units. Although there is nothing wrong in doing so, we prefer to distinguish between $[{\cal O}_i]_w$ from $[{\cal O}_i]$ and use the usual natural units.}  In other words, a theory is renormalizable if all the operators ${\cal O}_i$ appearing in the Lagrangian have weighted scaling dimensions $ [{\cal O}_i]_w$  (not to be confused with the standard scaling dimensions $[{\cal O}_i]$) which are not greater than $z+D$. 
Thus, the second term in (\ref{Lquad}), although manifestly irrelevant in the standard sense, behaves (and should be considered) as a marginal operator in this theory. 

It is useful to illustrate this construction with a specific simple example, namely a scalar field in $3+1$  space-time dimensions ($D=3$) and $z=2$. For simplicity, we also impose a $\Z_2$ discrete symmetry $\phi\rightarrow -\phi$. The most general renormalizable Lagrangian, invariant under the transformations (\ref{scalingW}),  is given by
\be
{\cal L}_r=\frac 12 \dot\phi^2-\frac{a^2}{2\Lambda^2} (\Delta \phi)^2 - \frac{h_2}{48 \Lambda^4} (\pa_i \phi)^2 \phi^4 -\frac{g_4}{10!\Lambda^6} \phi^{10}\ ,
\label{lere}
\ee
where $\Delta = \pa_i \pa_i$ is the Laplace operator in the spatial directions. All the operators appearing in (\ref{lere}) are weighted marginal.
The renormalizability properties of the theory are not changed if the scaling symmetry (\ref{scalingW}) is softly broken 
by adding weighted relevant operators. They are given by 
\be
{\cal L}_{\rm sr}=-\frac{m^2}{2}\phi^2-\frac{c^2}2 (\pa_i\phi)^2+\sum_{n=1}^3 \frac{g_n}{(2n+2)! \Lambda^{2(n-1)}} \phi^{2n+2} +\frac{h_1}{4\Lambda^2} (\pa_i \phi)^2\phi^2\,, 
\label{Vsr}
\ee
so that the final Lagrangian is the sum of eqs.(\ref{lere}) and (\ref{Vsr}).
In conventional scalar field theories in $3+1$ dimensions,  the interactions appearing in (\ref{lere})  would be non-renormalizable. What renders the theory renormalizable -- and the interactions in (\ref{lere}) weighted marginal -- is the modification of the propagator, which in momentum space now reads
\be
i\Big(k_0^2- c^2 k^2 -\frac{a^2}{\Lambda^2} k^4-m^2\Big)^{-1}\ ,
\label{aqui}
\ee
with $k^2=k_i k_i$. The $1/k^4$ high-energy behavior of the propagator 
leads to an improvement of the ultraviolet behavior of the theory. As a result, if no coupling of higher dimension is added,
 the theory is power-counting renormalizable. In the UV  the RG evolution of the weighted relevant parameters, such as $c^2$ or $g_1$ in eq.(\ref{Vsr}), will be governed by the RG evolution of a combination of the weighted marginal couplings $h_2$, $g_4$ and $a^2$. 
As we will see, in the IR the Lifshitz-type theory turns into a low-energy effective theory, where the weighted marginal couplings turn back into standard irrelevant ones and do not effectively run anymore. 
In this regime, if no Lorentz violating parameter appears in the Lagrangian terms with standard dimension $\leq 4$, then we effectively
recover Lorentz symmetry in the IR, which protects $c^2$ from any possible running. On the other hand, if some Lorentz violating
parameter is left (like e.g. $\delta c =c_{\phi_1}-c_{\phi_2}\neq 0$ in a two-field model), it will still generically induce a running of $c^2$, governed now by standard marginal couplings.

{}For the 3+1 dimensional model defined by eqs. (\ref{lere}) and (\ref{Vsr}), there is no renormalization of the couplings at one-loop level, due to the fact that the vertices in (\ref{lere}) involve at least six fields. For this reason, in what follows we will consider higher dimensional
scalar field theories, for which the weighted marginal vertices contain less fields and, as we will see, 
there is a non-trivial renormalization of the couplings already at one-loop level.

\section{UV behavior: a model with $z=2,~D=4$}

\subsection{One scalar particle}

We look for a weighted renormalizable scalar field theory with a non-trivial renormalization of the 
$(\pa_i \phi)^2$ operator at one-loop level. A simple quantum field theory of this sort is obtained in $4+1$ space-time dimensions with anisotropic scaling $z=2$.  
The most general renormalizable Lagrangian, up to total derivative terms and including all possible weighted marginal and relevant operators, is
\be
{\cal L}=\ha \dot\phi^2-\frac{a^2}{2\Lambda^2} (\Delta \phi)^2-\frac{c^2}{2}(\pa_i\phi)^2-\frac{m^2}{2} \phi^2 - \frac{\lambda}{4! \Lambda} \phi^4 -\frac{g}{4\Lambda^3} (\phi \pa_i \phi)^2-\frac{k}{6!\Lambda^4} \phi^6\,.
\label{LsingleD4}
\ee
All couplings appearing in ${\cal L}$, but the mass $m$ and $\Lambda$, are dimensionless. In order to reduce the number of operators,  we have imposed on ${\cal L}$ a discrete $\Z_2$ symmetry under which $\phi\rightarrow - \phi$. For simplicity, in the following we set $\Lambda = 1$.

Our first aim is to renormalize the theory at one-loop level and study the RG flows of the weighted marginal couplings $g$ and $k$ in the UV. The coupling $a^2$, although weighted marginal, is one-loop finite, since there is no way to extract four powers of external spatial momentum from the tadpole graph given by the quartic couplings appearing in ${\cal L}$. Similarly, the wave function renormalization of $\phi$ is trivial at one-loop level, $Z=1+{\cal O}($2-loops). Once  the RG flows for $g$ and $k$ are solved, we will  study the evolution of the weighted relevant coupling $c^2$.\footnote{Strictly speaking, the RG evolution of  $c^2$, as determined in a physical scheme, starts at two-loop order, even in presence of the quartic derivative interaction, since the momentum-dependence of the one-loop graph (which is a tadpole) is trivial. However, we can still define a running $c^2$ coupling by adding a fictitious momentum in the loop, seen as the momentum carried
by the composite operator $(\partial_i \phi)^2$. This is a standard trick. See e.g. \cite{Peskin} for the completely analogous case 
of the one-loop RG evolution of the mass parameter in the usual $\phi^4$ theory.}
We regularize the theory using a variant of dimensional regularization applied only to the spatial directions ($D=4-\epsilon$) and thus renormalize using a minimal subtraction scheme where only the poles in $1/\epsilon$ are subtracted, with no finite term. 

\begin{figure}[h]
\begin{center}
\begin{picture}(300,130)(0,0)
\setlength{\unitlength}{.65pt}
\SetScale{.65}
\GCirc(100,100){40}{1}
\Vertex(100,60){2.8}
\Vertex(65.4,120){2.8}
\Vertex(134.64,120){2.8}
\Line(75,16.7)(100,60)
\Line(100,60)(125,16.7)
\Line(159.64,163.3)(134.64,120)
\Line(184.64,120)(134.64,120)

\Line(40.36,163.3)(65.36,120)
\Line(15.36,120)(65.36,120)

\Text(100,75)[]{$g$}
\Text(125,113)[]{$g$}
\Text(75,113)[]{$g$}

\GCirc(350,100){40}{1}

\Vertex(390,100){2.8}
\Line(419.39,140.45)(390,100)
\Line(419.39,59.55)(390,100)
\Line(437.55,115.45)(390,100)
\Line(437.55,84.55)(390,100)
\Text(380,100)[]{$k$}

\Vertex(310,100){2.8}
\Line(266.7,75)(310,100)
\Line(266.7,125)(310,100)
\Text(320,100)[]{$g$}
\Text(100,180)[]{$(a)$}
\Text(350,180)[]{$(b)$}
\end{picture}
\caption{One-loop graphs contributing to the renormalization of the $\phi^6$ vertex. All external momenta are vanishing.}
\label{1loopgraph}
\end{center}
\end{figure}
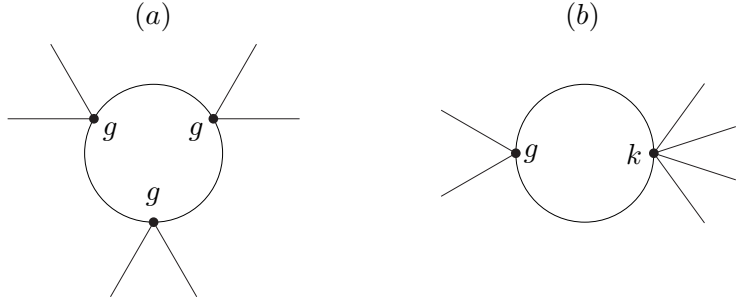
The superficial degree of divergence of a graph is easily computed by looking at the weighted scaling
of a graph. The one-loop corrections to the coupling $k$ come from a one-loop graph with 3 insertions of the coupling $g$ and from another graph with one insertion of $k$ and $g$. See figure \ref{1loopgraph}. Due to some unusual properties of these theories, we report, in  detail, the computation of the divergence term of the graph $(a)$ in fig. \ref{1loopgraph}. 
A divergence can only arise when all the momenta of the vertex are taken in the internal lines, so that  we can set to zero all external momenta $p_i$:
\be
(a) = (-i g\mu^\epsilon)^3 \frac{15}2 \int\frac{dq^0d^Dq}{(2\pi)^{D+1}} q^6 G(q^0,q^i)^3\ ,
\ee
where $15/2$ is a geometrical factor taking into account all possible channels and
\be
G(q^0,q^i) = \frac{i}{q_0^2 - a^2 q^4-c^2 q^2 -m^2}\ 
\ee
is the propagator for $\phi$. Here and in the following $q^2 \equiv q_i q_i$. 
After Wick rotating $q^0$ ($q^0=iq^5$), we can rewrite $(a)$ as
\bea
(a) &=&  -\frac{15 i g^3\mu^{3\epsilon}}4 \frac{d^2}{(dm^2)^2} \int_0^\infty\! \!d\alpha  \!\int\!\!\frac{dq_5d^Dq}{(2\pi)^{D+1}} q^6 e^{-\alpha(q_5^2 + a^2 q^4+c^2 q^2 +m^2)} 
\nn\\
&=&  -\frac{15 i g^3}4  I_{3,1} + {\rm finite} \ ,
\label{phi6div}
\eea
where
\be
I_{n,j}\equiv   \int_0^\infty \!d\alpha \, \alpha^{j+1} \!\int\frac{dq_5d^Dq}{(2\pi)^{D+1}} \, q^{2n} \, e^{-\alpha(q_5^2 + a^2 q^4+m^2)}\,.
\label{IntNM}
\ee
In writing the last equality of eq. (\ref{phi6div}), we have expanded the $\alpha c^2 q^2$ term in 
the exponential, since insertions of these terms lower the divergence of the graph. It is straightforward to check that the divergence arises only from the leading, $c^2$--independent term.
The integral (\ref{IntNM}) is easily done by going to radial coordinates and changing variables
$\alpha a^2 q^4 = r$. Performing the integrals we get
\be
I_{n,j} = \frac{\Omega_D \sqrt{\pi}}{4(2\pi)^{D+1}} (m^2)^{\frac{D-6+2n-4j}{4}}(a^2)^{-\frac{D+2n}{4}}
\Gamma\Big(\frac{D+2n}4\Big) \Gamma\Big(\frac{6+4j-D-2n}4\Big)\,,
\ee
where $\Omega_D = 2\pi^{D/2}/\Gamma(D/2)$ is the area of the $S^D$ sphere.
Using the same techniques, we can compute the graph $(b)$ as well.
By denoting $- \delta_k \phi^6/6!$ the Lagrangian counterterm canceling the divergences coming from the graphs $(a)+(b)$, we then get
\be
\delta_k = \Big(-\frac{45 g^3 l_4}{32a^5}+\frac{15 g k l_4}{8 a^3}\Big) \frac 1\epsilon\ ,
\label{deltak}
\ee
where we find convenient to express the result in terms of the usual loop factor for Lorentz invariant theories in $D$ space-time dimensions,  $l_D \equiv \Omega_D/(2\pi)^D$.  Similar manipulations allow to compute $\delta_g$, the coefficient of the counterterm  $- \delta_g (\phi\pa_i \phi)^2/4$:
\be
\delta_g= \frac{3 g^2 l_4}{8a^3} \frac 1\epsilon\,.
\label{deltag}
\ee
{}From  eqs.(\ref{deltak}) and (\ref{deltag}) we obtain the one-loop $\beta$ functions for $g$ and $k$:
\bea
\dot k \a = \a \beta_k =\frac{15l_4}8\frac{gk}{a^3} - \frac{45l_4}{32} \frac{g^3}{a^5} \,, \nn \\
\dot g \a = \a \beta_g = \frac{3l_4}8 \frac{g^2}{a^3}\,,
\label{RGphi}
\eea
where a dot stands for a derivative with respect to $t=\ln \mu/\mu_0$ and $\mu_0\gg 1$ is a given UV reference scale. Note that the effective couplings of the theory are 
\be
\hat g \equiv \frac{g}{a^3} \,, \ \ \ \ \ \ \hat k \equiv \frac{k}{a^4}\,.
\ee
The solutions of the RG equations (\ref{RGphi}) are 
\bea
\hat g(t)\a  = \a \frac{\hat g_0}{1-\frac{3l_4  \hat g_0}{8} t}\,, \nn \\
\hat k(t) \a = \a \hat k_0 \bigg(\frac{\hat g(t)}{\hat g_0}\bigg)^{5}
+\frac{5\hat g_0^2}{4} \bigg(\frac{\hat g(t)}{\hat g_0}\bigg)^{2}
\Bigg[1-\bigg(\frac{\hat g(t)}{\hat g_0}\bigg)^{3}\Bigg] \,,
\label{gk}
\eea
with $\hat g_0 = \hat g(0)$, $\hat k_0= \hat k(0)$. 
Since a Landau pole appears at the scale
\be
E_{pole} = \mu_0 e^{\frac{8}{3l_4 \hat g_0}}\,,
\ee
the range in which eqs.(\ref{gk}) are reliable is $1\ll E\ll E_{pole}$.

Having found the RG evolution of the weighted marginal couplings $k$ and $g$, we can now go on and study
the evolution of the weighted relevant coupling $c^2$. Its $\beta$ function reads
\be
{dc^2\over dt} = \beta_{c^2} = \frac{l_4 \hat g}{8} c^2\,,
\ee
giving
\be
c^2(t) = c^2_0 \bigg(\frac{\hat g(t)}{\hat g_0}\bigg)^{\frac 13}\,.
\label{Effc}
\ee 
Equation (\ref{Effc}) shows that in the UV regime $c^2$ has a logarithmic RG running, governed by the coupling $g$.  

We expect that in any generic, weakly-coupled quantum field theory of Lifshitz type, including also theories with gauge fields and matter in $D=3$, the  running of $c^2$ will be qualitatively similar to (\ref{Effc}), i.e. with a logarithmic dependence on the energy in the UV regime.

\subsection{Two scalar particles}

We will now show that theories with anisotropic scalings generically lead to different ``speed of light'' parameters (defined as coefficients of $(\vec\partial \phi_i )^2$) associated with different particles. More precisely, we will show
that the RG evolution of the difference $\delta c^2 \equiv c_1^2-c_2^2$, even in the most optimistic case
when $\delta c^2=0$ is an attractive fixed point, is generally too slow to give $\delta c^2 \simeq 0$ with the needed accuracy. A severe fine-tuning in the UV for $\delta c^2$ seems to be inevitable.

In what  follows we consider an  extension  of the single field model defined
by the Lagrangian density (\ref{LsingleD4})  to two fields $\phi_1$ and $\phi_2$, imposing 
the $\Z_2\times \Z_2$ symmetry $\phi_i\rightarrow -\phi_i$, $i=1,2$. The Lagrangian is given by
\be
{\cal L}_{2\phi} = {\cal L}_1+{\cal L}_2 - g_{12} (\phi_1 \pa_i \phi_1)(\phi_2 \pa_i \phi_2) - \frac{h_1}4 (\pa_i \phi_1)^2 \phi_2^2- \frac{h_2}4 (\pa_i \phi_2)^2 \phi_1^2- V_{12}(\phi_1,\phi_2)\,,
\label{L2fields}
\ee
where ${\cal L}_{1,2}$ are two copies of the Lagrangian appearing in (\ref{LsingleD4}) for the fields $\phi_1$ and $\phi_2$, and $V_{12}(\phi_1,\phi_2)$ is an additional potential:
\be
V_{12}(\phi_1,\phi_2) = \frac{\lambda_{12}}4 \phi_1^2 \phi_2^2 + \frac{k_{12}}{4!2} \phi_1^4 \phi_2^2 +  \frac{k_{21}}{4!2} \phi_2^4 \phi_1^2 \,.
\label{v12}
\ee
As can be seen, the Lagrangian contains a number of new interactions, which considerably complicate the analysis.
In particular, the one-loop renormalization of the couplings $k_1$, $k_2$, $k_{12}$ and $k_{21}$
involves diagrams with all possible combinations of three insertions of the quartic  couplings $g_{1,2}$,
$h_{1,2}$ and $g_{12}$ as well as diagrams with one insertion of any of the order six terms and one insertion of any of the quartic couplings. 
Fortunately, at one-loop level, as in the single field model considered before, the renormalization of $g_{1,2}$, $h_{1,2}$, $g_{12}$, $c_1^2$ and $c_2^2$ does not involve the $k_1$, $k_2$, $k_{12}$, $k_{21}$ couplings and therefore we do not need to compute the associated Feynman diagrams.
In analogy to eq. (\ref{gk}), there will always be a choice of boundary conditions for the couplings at $t=0$ such that the model is stable all the way down to the far UV.

Taking $a_1^2=a_2^2\equiv a^2$ for simplicity, after a straightforward but lengthy computation, we get 
\bea
\beta_{g_1}&  = &  \frac{l_4}{8} \Big(3 g_1^2 + 4 g_{12} h_2 + h_1 h_2 - 2 h_2^2\Big)\,,  \nn \\
\beta_{g_2}&  = &  \frac{l_4}{8} \Big(3 g_2^2 + 4 g_{12} h_1 + h_1 h_2 - 2 h_1^2\Big)\,,  \nn \\
\beta_{h_i}&  = &   \frac{l_4}{8} \Big(g_{12}^2 + h_i (g_1+g_2)+h_i^2 + 2 g_{12} h _i \Big)\,, \ \ \ \ i = 1,2 \ ,\nn \\
\beta_{g_{12}}&  = &   \frac{l_4}{16} \bigg[3 g_{12}^2 + 2g_{12}(g_1+g_2) +3 g_{12} (h_1+h_2) - h_1 h_2 \bigg]\,,
\label{RG2phi}
\eea
where all  couplings have been rescaled by a factor $1/a^3$ while keeping the same notation for the couplings for simplicity 
(i.e. we omit hats). The $\beta$ functions for the $c_i^2$ couplings are easily computed to be
\bea
\beta_{c^2_1}&  = &  \frac{l_4}{8} \Big(c_1^2 g_1 + c_2^2 h_1\Big) \nn \, ,\\
\beta_{c^2_2}&  = &  \frac{l_4}{8} \Big(c_2^2 g_2 + c_1^2 h_2\Big) \,.
\label{RG2c}
\eea
The RG equations (\ref{RG2phi}) do not admit a simple analytic solution in general.
A class of exact solutions  is however obtained by substituting the ansatz
\bea
&& g_1(t)=g_2(t) = g(t)=\frac{g_0}{ 1- x l_4 t}\,, \ \ \ 
\nn\\
&& h_1(t)=h_2(t) = h(t)=\frac{h_0}{ 1- x l_4 t}\,, \ \ \ 
\nn\\
&& g_{12}(t)=\frac{g_{12,0}}{ 1- x l_4 t}\ ,
\label{clases}
\eea
in eqs.(\ref{RG2phi}) and solving the (now algebraic) equations for $g_0,h_0,g_{12,0}$ and $x$:
 \bea
 8 x g_0-(3 g_0^2+4 g_{12,0} h_0-h_0^2) \a =\a 0  \,, \nn  \\ \nonumber
 8 x h_0-[g_{12,0}^2+2 g_{12,0} h_0 +h_0 (2 g_0+h_0)]\a =\a 0\,, \\
 16 x g_{12,0}-[3 g_{12,0}^2+g_{12,0} (4 g_0+6 h_0)-h_0^2]\a =\a 0\,.
\label{algebraic}
 \eea 
In terms of $g_0$, $h_0$ and $x$, the running for $c^2=(c_1^2+c_2^2)/2$ and $\delta c^2=(c_1^2-c_2^2)/2$ is given by
\be
c^2(t) = c^2_0 \bigg(\frac{g(t)}{g_0}\bigg)^{\frac{g_0+h_0}{8x}} \,, \ \ \ \ 
\delta c^2(t) = \delta c^2_0  \bigg(\frac{g(t)}{g_0}\bigg)^{\frac{g_0-h_0}{8x}}\,.
\ee
The system (\ref{algebraic}) is under-constrained (3 equations for 4 variables), so we fix one of the couplings, say $g_0=1$, and look for solutions for $x$ and for the other couplings $h_0,g_{12,0}$.
Taking $g_0=r$ would simply rescale the solutions $x\to r x,~ h_0\to r h_0,~g_{12,0}\to r g_{12,0}$, so that the RG evolution of $c^2$ and $\delta c^2$ is unaffected.
A sufficient condition to get a (semi)positive definite interaction requires $g_0, h_0 \geq 0$ and
$|g_{12,0}|\leq (g_0+h_0)/2$. We find seven solutions, one of which is unstable and is disregarded.
The remaining six solutions are  
\bea
1)\; x \a = \a \frac 38\,,  \hspace{2.7cm}  g_{12,0} =0\,, \hspace{2.1cm}  h_0=0\,; \nn \\
2)\; x \a = \a \frac 34\,,  \hspace{2.7cm}  g_{12,0} =1\,, \hspace{2.1cm}  h_0=1\,; \nn \\
3) \; x\a = \a \frac{5}{12}\,,  \hspace{2.5cm} g_{12,0} =\frac 13\,, \hspace{2cm}h_0=\frac 13\,; \nn \\
4-5) \; x \a = \a \frac{5}{16}(5\mp \sqrt{17}), \  \ \ \ \  \ \ g_{12,0} =\frac 12 (3\mp\sqrt{17})\,, \ \ \ h_0=\frac 12 (13\mp3\sqrt{17}) \nn \,; \\
6) \; x \a = \a \frac{1}{256}(77+5\sqrt{17}), \ \ g_{12,0} =\frac{1}{16} (7-\sqrt{17})\,, \ \ h_0=\frac 18 (1+\sqrt{17}) \,.
\eea
As can be seen from $x$, all six solutions correspond to couplings (and $c^2$) which grow in the UV in all cases. 
The deviation $\delta c^2$, instead, increases in the UV for 1), 3), 4) and 6),
is constant in the case 2) and decreases in the UV in 5).
Note that solution 1) reproduces the RG evolution (\ref{gk}) for a single field. 

We can also perturb the solutions found above and study the RG evolution of the
fluctuations at the linear level. We focus on the case 1), since it is the only one giving rise
to simple analytical results.  We look at linear perturbations around the solutions putting 
 \bea
 &&g_1=g+\delta g+\delta u\ , \qquad g_2=g-\delta g+\delta u \ , 
\nn\\
&& h_1=h+\delta h+\delta v\ , \qquad h_2=h-\delta h+\delta v \ , \qquad g_{12}\to g_{12}+\delta g_{12}\ ,
 \eea
and consider  $\delta c^2$ as a fluctuation. In this way, we get
\bea
\frac{\delta g(t)}{\delta g_0} \a =\a \frac{\delta u(t)}{\delta u_0} = \bigg(\frac{g(t)}{g_0}\bigg)^{2}\,,  \nn \\
\frac{\delta h(t)}{\delta h_0} \a = \a \frac{\delta v(t)}{\delta v_0} = \frac{\delta g_{12}(t)}{\delta g_{12,0}} = \bigg(\frac{g(t)}{g_0}\bigg)^{\frac 23} \,. \nn \\
\delta c^2(t) \a = \a \delta c^2_0  \bigg(\frac{g(t)}{g_0}\bigg)^{\frac 13}+ \frac{c^2_0 l_4 t}8
 \bigg(\frac{g(t)}{g_0}\bigg)^{\frac 43} \delta g_0+c_0^2\bigg[ \bigg(\frac{g(t)}{g_0}\bigg)^{\frac 13}-1\bigg] \delta h_0  \,.
 \label{deltaRG}
 \eea
 {}From eqs.(\ref{deltaRG}) we see that the fixed-point is stable, with all fluctuations decreasing towards the IR.  A similar study can be done for the other solutions, which are not all IR stable. From eq.(\ref{deltaRG})
we see that it is not enough to start  with $\delta c^2_0=0$ at $\mu_0\gg 1$ to ensure that $\delta c^2=0$ near $E\sim\Lambda$.  
In addition, one has to ensure that also other perturbations around some fixed point 
are fine-tuned to zero.

\begin{figure}[t!]
\begin{minipage}[t]{0.465\linewidth} 
\begin{center}
\includegraphics*[width=\textwidth]{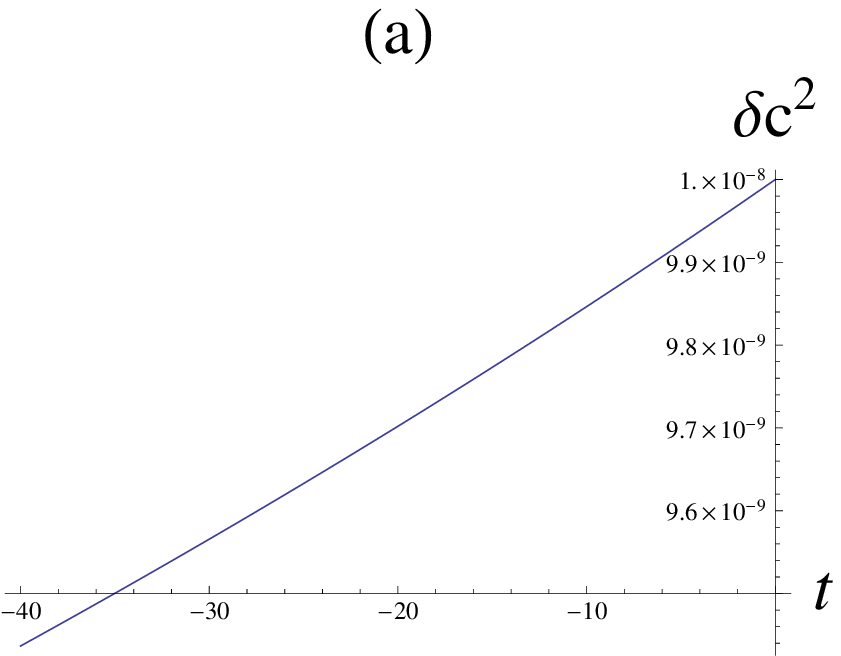}
\end{center}
\end{minipage}
\hspace{0.5cm} 
\begin{minipage}[t]{0.48\linewidth}
\begin{center}
\includegraphics*[width=\textwidth]{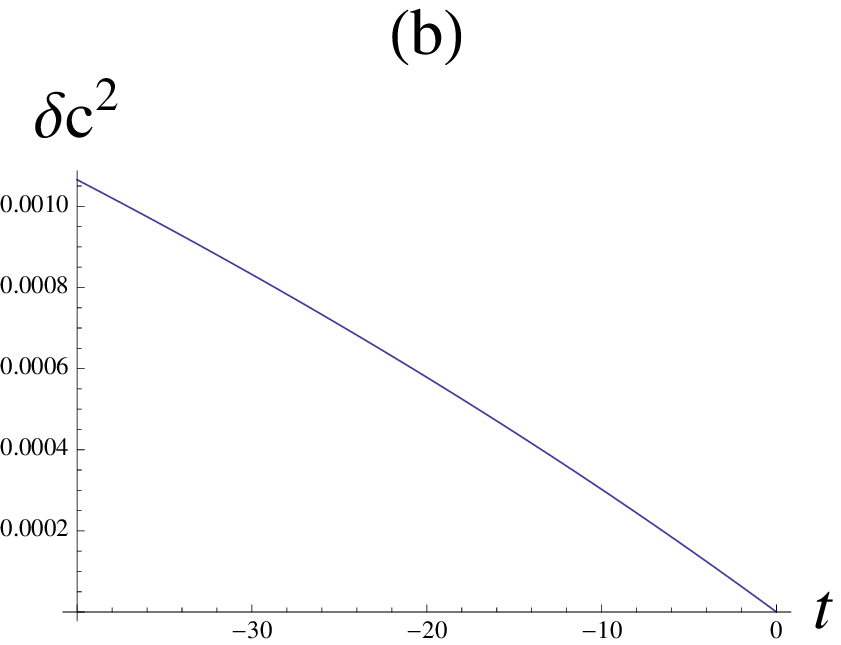}
\end{center}
\end{minipage}
\caption{(a) RG evolution of $\delta c^2$ as given by eq.(\ref{deltaRG}) for $c_0^2=1$, $\delta g_0= \delta h_0 = 0$, $\delta c_0^2=10^{-8}$. (b) RG evolution of $\delta c^2$ as given by eq.(\ref{deltaRG}) for $c_0^2=1$, $\delta g_0= \delta h_0 = -10^{-2}$, $\delta c_0^2=0$. }
\label{figdeltac}
\end{figure}

In section 5 we will argue that in the IR  the RG evolution of $\delta c^2$, as given by the contributions of only weighted marginal couplings, essentially stops.  Given the experimental bounds on $\delta c^2$ for ordinary particles mentioned before,
recovering Lorentz invariance in the IR with enough accuracy requires fine-tuning of 
$\delta c^2$, $\delta g$ and $\delta h$ in the UV to extremely small values. 
We illustrate this point in figures \ref{figdeltac}(a) and \ref{figdeltac}(b), where the running of $\delta c^2(t)$ over 40 orders of magnitude for an arbitrary given choice of boundary conditions is shown.

\subsection{Case of particle with no self-interactions}

The two-scalar model can be adapted to the case where one of the particles, say $\phi_1$, has no self-interaction terms, i.e. $g_1=0$.
This  case represents a  situation of physical interest: in electrodynamics the photon has no self-interaction term and one may wonder if the speed of light will still significantly depend on the energy.
Here we will show that the  energy dependence of either speed-of-light parameters $c_1$ or $c_2$ does not rely on the presence or absence of self-interaction terms.

Consider the beta functions (\ref{RG2phi}). An exact solution can be found by setting
\be
g_1=h_2=g_{12}=0\ .
\ee
We are left with two equations
\be
\dot h_1 = \frac{l_4}{8} h_1 p\ ,\qquad \dot p = \frac{l_4}{8}  (3p^2- 5p h_1+ h_1^2)\ ,\qquad \ p\equiv g_2+h_1  \,.
\label{h1peqs}
\ee
Next, we write $p= f h_1$, so that the second equation takes the form
\be
\dot f = \frac{l_4}{4} h_1 (f-f_+)(f-f_-)\ , \qquad f_\pm ={1\over 4}(5\pm\sqrt{17})\ .
\label{hny}
\ee
One obvious solution is $\dot f=0$ which requires $f=f_+$ or $f=f_-$. 
The solution with $f=f_-$ is not physical, since $f=1+g_2/h_1>1$.
Setting $f=f_+$  leads to a solution similar to
the class of solutions considered in (\ref{clases}),
\be
g_2(t)= {g_0\over 1-x l_4 t}\ ,\qquad h_1(t)= {h_0\over 1-x l_4 t}\ ,\qquad x= {f_+ h_0\over 8}\ .
\label{consa}
\ee
Now consider solutions where $f$ is not constant.
Writing
\be
\dot f ={df\over dh_1}\dot h_1 = {df\over dh_1}  \frac{l_4}{8} h_1^2 f\,,
\ee
eq. (\ref{hny}) becomes
\be
{f\over (f-f_+)(f-f_-)}\ {df\over dh_1}={2\over h_1}\ .
\ee
We can now integrate this equation and obtain
\be
h_1 = k_0\  {(f-f_+)^{f_+\over 2(f_+-f_-)}\over (f-f_-)^{f_-\over 2(f_+-f_-)}} \,,
\label{h1-f}
\ee
where $k_0$ is an integration constant. Inserting eq.(\ref{h1-f}) into the first equation in (\ref{h1peqs})
gives a first order differential equation for $f$ that can be integrated. The result is
\be
\frac{l_4 k_0}4  \ t= -\int_{f}^\infty \ {df^\prime\over  (f^\prime-f_+)^{1 + {f_+\over 2(f_+-f_-)}} (f^\prime-f_-)
^{1-{f_-\over 2(f_+-f_-)}}    } \,.
\label{solfin}
\ee
We have chosen $\mu_0$ as the ultra-high energy scale at which $f\to \infty·$, so that $f$ is defined in the interval $-\infty<t <0$. The integral (\ref{solfin}) defines $f=f(t)$ and hence $h_1(t)=h_1[f(t)]$ and $ g_2(t)= (f(t)-1)h_1(t)$. It can be expressed in terms of hypergeometric functions,
but its explicit expression is not needed in order to see the qualitative 
behavior of the couplings in the regime $t\ll 0$.
In this limit,  $f\to f_+ $  and hence one approaches the constant $f$ solution (\ref{consa}) 
discussed above. For large $(-t)$, one has 
the behavior, 
\be
(f-f_+)^{f_+\over 2(f_+-f_-)} \sim h_1 \sim g_2 \sim  {1\over |t|}\,.
\label{h1g2IR}
\ee
Hence both $h_1$ and $g_2$ decrease towards the IR. 

Next, we consider the RG equations (\ref{RG2c}) for $c_1$ and $c_2$, which simplify to
\be
{d c^2_1\over dt}  = c_2^2 h_1\ ,\qquad\ {d c_2^2\over dt}= c_2^2 g_2\ .
\ee
{} Integrating these equations  using the behavior of $g_2$ and $h_1$ given in eq.(\ref{h1g2IR}),  one can see that
$c_2$ goes to zero like $1/|t|^k$, with $k>0$, while $c_1^2$ approaches a constant value
for $t\ll 0$.

\section{UV behavior: a UV free model with $z=2$, $D=10$}

In this section we look for a weighted renormalizable scalar field theory which is UV free.
In order to simplify our analysis, we consider a $\phi^3$ model which is weighted-counting renormalizable in $D=10$ spatial dimensions. Although the $\phi^3$ model is manifestly an unphysical
theory, having an unbounded potential, this instability does not affect the RG equations, which can then be formally studied and hopefully seen as a toy laboratory for more complicated stable UV free theories, such as Yang-Mills (YM) theories. As in section 3, we first study the one-loop RG evolution of a single field, which will enable us to compute the RG evolution of $c^2$. Then we shall consider a two-field model, where $\delta c^2$ will also be considered.

\subsection{One scalar}

The most general renormalizable Lagrangian reads
\be
{\cal L}=\ha \dot\phi^2-\frac{a^2}{2} (\Delta \phi)^2-\frac{c^2}{2}(\pa_i\phi)^2-\frac{m^2}{2} \phi^2 - \frac{\lambda}{3!}  \phi^3 \,.
\label{Lsingle}
\ee
As in section 2, we use a dimensional regularization in the spatial directions only. We will add counterterms to  subtract the poles in $1/\epsilon$ only, with no finite term. 
We study the theory in the energy range $E\gg 1$, where the analysis is reliable.

The wave-function renormalization in the $\phi^3$ theory is non-trivial and hence the relevant one-loop graphs to compute are those associated with the two and three point functions. These graphs are computed exactly along the same lines followed in section 3, so we will just report the results for the $\beta$ functions and the anomalous dimension $\gamma$ of $\phi$. We find $\gamma=l_{10}\lambda^2/(64 a^5)$ and
\be
\beta_{a^2} = 2 \omega_a \frac{\lambda^2}{a^5} a^2 \,, \ \ \ \ 
\beta_\lambda  = -\omega_\lambda\frac{\lambda^2}{a^5}\lambda \,,
\label{RGphi3}
\ee
where $\omega_\lambda = 9l_{10}/64$, $\omega_a =21l_{10}/640$.  
The effective coupling of the theory is
\be
x= \frac{\lambda^2}{a^{5}}\,.
\ee
The RG equations are easily solved in terms of $x$. One has
\be
\dot{x} = - \omega_x x^2\,,
\ee
with $\omega_x=2\omega_\lambda+5 \omega_a$, and thus
\be
x(t) = \frac{x_0}{1+x_0 \omega_x t}\,.
\label{xsol}
\ee
Plugging eq.(\ref{xsol}) into eqs.(\ref{RGphi3}) give
\be
\lambda(t) = \lambda_0\bigg(\frac{x(t)}{x_0}\bigg)^{\frac{\omega_\lambda}{\omega_x}}\,, \ \ \ 
a^2(t) = a^2_0\bigg(\frac{x(t)}{x_0}\bigg)^{-\frac{2\omega_a}{\omega_x}} \,.
\label{RGsolphi3}
\ee
The coupling $a^2$ increases in the UV, while $\lambda$ and the effective coupling $x$ are UV free.

Let us now study the evolution of $c^2$. Its associated $\beta$ is
\be
\beta_{c^2} = \omega_c x c^2 \,,
\label{betacphi3}
\ee
where $\omega_c = 3l_{10}/16$. Hence 
\be
c^2(t) = c^2_0\bigg(\frac{x(t)}{x_0}\bigg)^{-\frac{\omega_c}{\omega_x}}\,.
\ee
The RG evolution of $c^2$ is again governed by the weighted marginal couplings of the theory.
Interestingly enough, $c^2$ logarithmically
increases in the UV, as in the model of section 3.

\subsection{Two scalars}

We can add a further scalar $\eta$ to the $\phi^3$ model and, for simplicity, we impose a discrete $\Z_2$ symmetry  under which $\eta\rightarrow - \eta$. The Lagrangian is given by
\bea
{\cal L}  \a =  \a \ha \dot\phi^2-\frac{a^2}{2} (\Delta \phi)^2-\frac{c^2}{2}(\pa_i\phi)^2-\frac{m^2}{2} \phi^2 - \frac{\lambda}{3!}  \phi^3 + \nn \\
\a\a  \frac 12 \dot \eta^2 - \frac{\tilde a^2}2 (\Delta  \eta)^2 - \frac{\tilde c^2}2 (\partial_i \eta)^2 - \frac{\tilde m^2}2  \eta^2  - \frac{\tilde \lambda}{2} \eta^2 \phi \,.
\eea
The computation of the $\beta$ functions for $\lambda$, $\tilde \lambda$, $a^2$, $\tilde a^2$, $c^2$ and $\tilde c^2$ is straightforward, although a bit laborious, mainly because we keep $a^2\neq \tilde a^2$ in general.
We find
\bea
\beta_\lambda & = &  \left[- \tilde \lambda^3 F(\tilde a,\tilde a)+\lambda ^3 \left(- F(a,a)+\frac{3}{2} K(a,a)\right)+\frac{3}{2} \lambda  \tilde \lambda^2 K(\tilde a,\tilde a)\right] \frac{l_{10}}{4} \,, \nn \\
\beta_{\tilde \lambda} & = &  
\left[- \lambda  \tilde \lambda^2 F(a,\tilde a)+\frac{1}{2} \lambda ^2 \tilde \lambda K(a,a)+\tilde \lambda^3 \left(- F(\tilde a,a)+2 K(a,\tilde a)+\frac{1}{2} K(\tilde a,\tilde a)\right)\right]\frac{l_{10}}4 \,, \nn \\
\beta_{a^2} & = & \left[\lambda ^2 \left( a^2 K(a,a)+Q(a,a)\right)+\tilde \lambda^2 \left( a^2 K(\tilde a,\tilde a)+ Q(\tilde a,\tilde a)\right)\right]\frac{ l_{10}}4 \,, \nn \\
\beta_{\tilde a^2} & = &  \tilde \lambda^2 \left(\tilde a^2 K(a,\tilde a)+Q(a,\tilde a)\right)\frac{ l_{10}}2 \,, \nn  \\
\beta_{c^2} & = & \left[\lambda ^2 \left(c^2 K(a,a)+R(a,a,c,c)\right)+\tilde \lambda^2 \left(c^2 K(\tilde a,\tilde a)+ R(\tilde a,\tilde a,\tilde c,\tilde c)\right)\right] \frac{l_{10}}{4}\,, \nn \\
\beta_{\tilde c^2} & = &  \tilde \lambda^2 \left(\tilde c^2 K(a,\tilde a)+R(a,\tilde a,c,\tilde c)\right) \frac{l_{10}}{2} \,,
\label{betas}
\eea
where
\bea
F(a_1,a_2) & = &   \frac{2 a_1+ a_2}{a_1^3 a_2 (a_1+ a_2)^2}\,, \nn \\
K(a_1,a_2) & = &  \frac{1}{a_1 a_2 (a_1+a_2)^3}\,, \nn \\
Q(a_1,a_2) &  = & \frac{(a_1^4+5 a_1^3 a_2+10 a_1^2 a_2^2+ 5 a_1 a_2^3+a_2^4)}{5 a_1 a_2 (a_1+a_2)^5}\,, \nn \\
R(a_1,a_2,c_1,c_2) &  = & \frac{c_1^2 a_2^2 (3 a_1^3+12 a_1^2 a_2+8 a_1 a_2^2+ 2 a_2^3)+(1\leftrightarrow 2)}{5 a_1^3 a_2^3 (a_1+a_2)^4}\,.
\eea
When $\tilde \lambda=0$, eqs.(\ref{betas}) collapse to eqs.(\ref{RGphi3}) 
and (\ref{betacphi3}), as expected.
Finding the general, analytic solutions to eqs.(\ref{betas})  is complicated.
Interestingly, eqs.(\ref{betas}) admit, nevertheless, a simple fixed-point solution given by
$\lambda = \tilde \lambda\equiv \bar \lambda(t)$, $a^2=\tilde a^2\equiv  \bar a^2(t)$, $c^2=\tilde c^2\equiv \bar c^2(t)$.
The RG equations for $\bar\lambda$, $\bar a^2$ and $\bar c^2$ are precisely given by eqs.(\ref{RGphi3}) 
and (\ref{betacphi3}), provided one makes the substitution $\omega_\lambda \rightarrow  2 \omega_\lambda$,
$\omega_a \rightarrow  2 \omega_a$, and $\omega_c \rightarrow  2 \omega_c$. For completeness, below we report  the corresponding solutions:
\be
\bar \lambda(t)  =  \bar \lambda_0\bigg(\frac{\bar x(t)}{\bar x_0}\bigg)^{\frac{\omega_\lambda}{\omega_x}}\,, \ \ \ 
\bar a^2(t) = \bar a^2_0\bigg(\frac{\bar x(t)}{\bar x_0}\bigg)^{-\frac{2\omega_a}{\omega_x}} \,, \ \ \ 
\bar c^2(t) = \bar c^2_0\bigg(\frac{\bar x(t)}{\bar x_0}\bigg)^{-\frac{\omega_c}{\omega_x}}\,.
\label{RGsolphi30}
\ee
where
\be
\bar x(t) \equiv \frac{\bar \lambda^2(t)}{\bar a^5(t)} = \frac{\bar x_0}{1+2\bar x_0 \omega_x t}\,.
\ee
The fixed-point solution (\ref{RGsolphi30}) is unstable under small deformations.
In order to show that, we consider the following ``symmetric" perturbations
\bea
\lambda & = &  \bar \lambda + \delta \lambda\,, \ \ \ \ \ \ \ \ \ \tilde \lambda =  \bar \lambda - \delta \lambda\,, \nn \\
a^2 & = &  \bar a^2+ \delta a^2 \,, \ \ \ \ \ \ \tilde a^2 = \bar a^2- \delta a^2\,, \nn \\
c^2 & = &  \bar c^2+ \delta c^2\,, \ \ \ \ \ \ \ \tilde c ^2= \bar c^2- \delta c^2 \,,
\eea
keeping up to linear terms in the perturbations.  We get
\bea
\beta_{\delta\lambda}  \a = \a  -2\omega_{\delta \lambda} \frac{\bar \lambda^2}{\bar a^5}  \delta \lambda\,, \nn \\
\beta_{\delta a^2} \a =\a \frac{l_{10}}{160}  \frac{\bar \lambda}{\bar a^5} (21 \bar a^2 \,\delta \lambda +10\delta a^2 \, \bar \lambda)\,, \ \ \  \nn \\
\beta_{\delta c^2} \a = \a \frac{l_{10}}{16}   \frac{\bar \lambda}{\bar a^5} (6 \bar c^2\, \delta \lambda + \delta c^2\,  \bar \lambda) \,,
\eea
where $\omega_{\delta\lambda} = 7l_{10}/64$. The UV evolution of the couplings is therefore given by 
\bea
\delta \lambda(t) & = & \delta \lambda_0\bigg(\frac{\bar x(t)}{\bar x_0}\bigg)^{\frac{\omega_{\delta\lambda}}{\omega_x}}\,, \nn \\
\delta a^2(t) &=& \Big(\delta a_0^2 -\frac{a_0^2\delta \lambda_0}{\bar \lambda_0}\Big)
\bigg(\frac{\bar x(t)}{\bar x_0}\bigg)^{-\frac{4}{57}}+{a_0^2\delta \lambda_0\over \bar\lambda_0} \ \bigg(\frac{\bar x(t)}{\bar x_0}\bigg)^{-\frac{62}{285}}\,,  \nn \\
\delta c^2(t) &=& \Big(\delta c_0^2 - \frac{c^2_0\delta \lambda_0}{\bar \lambda_0}\Big)
\bigg(\frac{\bar x(t)}{\bar x_0}\bigg)^{-\frac{4}{57}} +{c_0^2\delta \lambda_0\over \bar\lambda_0} \bigg(\frac{\bar x(t)}{\bar x_0}\bigg)^{-\frac{28}{57}}\,.
\label{deltacphi3}
\eea
The fixed-point is stable in the UV only along $\delta \lambda$.
The situation is similar to the one encountered in the model of section 3.
The RG evolution of $\delta c^2$ in the UV does not help in alleviating the fine-tuning needed to get   $\delta c^2$ small enough for energies near $\Lambda$.

We expect that the general lessons for the behavior of the different couplings under the Renormalization Group learned in this section
 should  be qualitatively shared by a large class of UV free quantum field theories, including the relevant case of perturbative YM theories coupled to matter. 

\section{IR behavior of Lifshitz-type theories}

As well known, in conventional Lorentz invariant theories, the MS scheme has to be used with care in studying the evolution of the couplings in presence of mass terms when
$E\ll m$, since the decoupling of massive particles is not manifest.  Indeed, the MS $\beta$-functions, being mass--independent, are formally the same for any $E$, 
while for $E\ll m$ the physical coupling does not effectively run anymore. 
In conventional theories, a reasonable approximation to overcome this problem and still use the MS scheme is to take an effective approach, where
for $E\ll m$ the heavy particle is integrated out and its contribution to the running neglected. 
One could also make a more refined study
of the transient region around $E\sim m$
(we will not do it in our similar case below).
Since we are investigating here unconventional theories of the Lifshitz type, 
it is instead preferable to use a more physical renormalization scheme, such as the momentum subtraction scheme, where the $\beta$-functions are mass dependent and the decoupling is manifest, even if the associated one-loop computations become necessarily more involved. 
In the present case, because the propagator is given by
\be
i\Big(k_0^2- c^2 k^2 -\frac{a^2}{\Lambda^2} k^4-m^2\Big)^{-1}\ ,
\label{aqua}
\ee
IR effects are expected as soon as $c^2\Lambda^2 k^2 \sim a^2k^4$, i.e. at momentum scales of order $c\Lambda/a$. The scale $c\Lambda /a$ plays now the role of $m$.
As explained below, the underlying reason is that the term $c^2 k^2 $  in the propagator itself provides an IR regularization of the amplitudes even when $m=0$.

\subsection{$\beta $-function in the IR in conventional $\phi^4$ theory}

Let us begin by briefly reviewing the standard computation of the one-loop $\beta$-function of the Lorentz-invariant $\phi^4$ theory in $3+1$ dimensions in the momentum subtraction scheme. Let $\Gamma^{(4)}$ be the tree+one-loop+counterterms 1PI four-point function. The physical coupling $\lambda$ can be defined as the value of $\Gamma^{(4)}$ at some given energy scale. In terms of the Mandelstam variables $s,t,u$, we can define
\be
\Gamma^{(4)}(s=t=u=-\mu^2) \equiv \Gamma^{(4)}(\mu) = - \lambda\,.
\label{MOM}
\ee
Here we used the standard trick of introducing  the renormalization group scale at some euclidean value of the kinematical invariants to circumvent threshold singularities.
Eq.(\ref{MOM}) fixes the value of the counterterms, so that 
\be
\Gamma^{(4)}(s,t,u) = -\lambda + \Gamma^{(4)}_l(s,t,u)- \Gamma^{(4)}_l(\mu)\,,
\label{MOM2}
\ee
where $\Gamma^{(4)}_l$ is purely the 1-loop contribution to $\Gamma^{(4)}$.
Since $\beta_{m^2}$ and the anomalous dimension of $\phi$ vanish at one-loop order, being $\Gamma^{(2)}_l$
momentum-independent, inserting eq.(\ref{MOM2}) in the Callan-Symanzik (CS) equation satisfied by $\Gamma^{(4)}$
gives directly $\beta_\lambda=\mu d \lambda/d\mu$:
\be
\beta_\lambda =   -  \mu\frac{\partial \Gamma_l^{(4)}}{\partial \mu} \,.
\label{MOM3}
\ee
Using standard techniques, we get
\be
\beta_\lambda =  6  \lambda^2 \int\frac{d^4k_E}{(2\pi)^4}\int_0^1\!\!dx  \frac{\mu^2 x(1-x)}{[k_E^2+m^2+\mu^2 x (1-x)]^{3}} =
\frac{3\lambda^2}{16\pi^2} \int_0^1 \!\! dx \frac{\mu^2 x(1-x)}{m^2+\mu^2 x(1-x)} \,.
\label{phi4MOM}
\ee 
The integral in $x$ in eq.(\ref{phi4MOM}) can analytically be performed, but it is not necessary to do so in order to see the 
UV and IR behavior of $\beta_\lambda$. For $\mu^2\gg m^2$, $\beta_\lambda \simeq  3\lambda^2/(16\pi^2)$, in agreement with what
one would have obtained with, say, a computation in the MS scheme. For $\mu^2\ll m^2$, $\beta_\lambda \simeq \lambda^2/(32\pi^2) (\mu^2/m^2)$.
The behavior of $\lambda$ below $m$, its freezing, can essentially be understood by noticing that $m$ acts as an IR regulator to the one-loop graph,
which would otherwise have an IR divergence when $\mu\rightarrow 0$. 

\subsection{Computation of $\beta$-functions of Lifshitz-type theories in the IR}

As we have already mentioned at the beginning of section 5, 
even in the absence of mass terms, the RG evolution of coupling constants presents at least two regimes, characterized by $\mu\gg c/a$ (UV) and $\mu\ll c/a$ (IR).\footnote{In units of $\Lambda$. In the following, $\Lambda$ will be restored in some formulas when convenient.}  Here we consider the $\phi^3$ model of section 4 in detail as an illustrative case.   As it will shortly be clear, 
the important qualitative aspects of the results should be completely general and apply to any Lifshitz-like theory, including the model in section 3.
We define the renormalized field $\phi$ and the couplings $\lambda$, $a^2$ and $c^2$ as follows:
\bea
\frac{\partial \Gamma^{(2)}}{\partial (p^0)^2}\bigg|_0 \a = \a 1\,; \ \ \  \frac 1{4!} \frac{\partial^4  \Gamma^{(2)}}{\partial p^4}\bigg|_0  = -a^2\,;
\ \ \ \frac 12 \frac{\partial^2  \Gamma^{(2)}}{\partial p^2}\bigg|_0  =  - c^2 \,, \nn \\
\Gamma^{(3)}[(p_1^0)^2\a =\a -\omega(\mu^2),p_{2,3}^0=\vec{p}_{1,2,3}=0] \equiv \Gamma^{(3)}(\mu) = - \lambda,
\eea
where $\Gamma^{(2)}$ and $\Gamma^{(3)}$ are the tree+one-loop+counterterms 1PI two- and three-point functions and $0$ stands for the subtraction point  $(p^0)^2=-\omega(\mu^2) = - a^2 \mu^4 - c^2 \mu^2$, $ p=|\vec{p}|=0$. We define the subtraction point at vanishing spatial momenta for technical reasons, since the computation is greatly simplified in this way. Due to the Lifshitz nature of the theory, the energy $p^0$ should depend quadratically (for $z=2$) on the sliding scale $\mu$, so in the UV we get $\omega(\mu^2)\simeq a^2 \mu^4$. The factor $a^2$ has been inserted just to slightly simplify the analysis that will follow, but it is by no means necessary.
In the IR we get $\omega(\mu^2)\simeq c^2 \mu^2$.  In both cases, we have chosen the subtraction point at  an euclidean energy scale to circumvent threshold singularities; this is a non-relativistic analog of the more familiar condition (\ref{MOM}).
Using the CS equations satisfied by $\Gamma^{(2)}$ and $\Gamma^{(3)}$, we can derive
$\beta_\lambda$,  $\beta_{a^2}$, $\beta_{c^2}$ and the anomalous dimension $\gamma$ of $\phi$ in terms of  the purely one-loop contributions (denoted by $\Gamma^{(2)}_l$ and $\Gamma^{(3)}_l$) to  $\Gamma^{(2)}$ and $\Gamma^{(3)}$. We have
\bea
\gamma \a= \a -\frac 12 \mu  \frac{\partial  \dot \Gamma^{(2)}_{l,0}}{\partial \mu} \,, \hspace{2.2cm}
\beta_\lambda = - \mu \frac{\partial \Gamma^{(3)}_{l,0}}{\partial\mu}+3 \lambda \gamma\,, \nn \\
\beta_{a^2}\a = \a -\frac{1}{4!} \mu \frac{\partial \Gamma^{(2)\prime\prime\prime\prime}_{l,0}}{\partial \mu} + 2\gamma a^2\,, \ \ \ 
\beta_{c^2} = -\frac{1}{2} \mu \frac{\partial \Gamma^{(2)\prime\prime}_{l,0}}{\partial \mu} + 2\gamma c^2\, . \ \ \
\label{gammaphi3} 
\eea
To simplify the notation, here we have denoted by a dot and a prime a derivative with respect to $(p^0)^2$ and $p$, respectively. 
We now show the computation of $\gamma$ in some detail. One has
\bea
\gamma \a = \a 3  \lambda^2 \int\frac{d^{10}k dk_E }{(2\pi)^{11}}\int_0^1\!\!dx  \frac{ \mu^2 \omega^\prime(\mu^2) x^2(1-x)^2}{[k_E^2+a^2k^4+c^2 k^2+\omega(\mu^2) x (1-x)]^{4}} \nn \\
\a = \a
\frac{15 l_{10}\lambda^2}{32} \int_0^1 \!\! dx \int_0^\infty \!\! dk  \frac{k^9 \mu^2 \omega^\prime(\mu^2) x^2(1-x)^2}{[a^2k^4+c^2 k^2+\omega(\mu^2) x (1-x)]^{7/2}}\,.
\label{MOMLif2}
\eea
Computing the integral in  $x$, we find
\be
\gamma = \lambda^2l_{10} \int_0^\infty \! dk \frac{k^8\mu^2\omega^\prime(\mu^2)}{\sqrt{a^2 k^2+c^2}[4(a^2k^4+c^2k^2) +\omega(\mu^2)]^3}\,.
\label{gammaphi}
\ee
Note that the integrand in eq.(\ref{gammaphi}) is UV {\it and} IR finite for any value of $\mu^2>0$. In the UV, $\mu\gg c \Lambda/a $,  the $c^2$ terms can be neglected, 
in which case the integral is easily performed giving
\be
\gamma^{\rm (UV)} \simeq \frac{l_{10}  \lambda^2}{64a^5}\,, \ \ \ \  \mu\gg {c \over a}\ \Lambda \,.
\ee
In the IR, one can neglect the $\mu^2$ term appearing in the denominator of the integrand of eq.(\ref{gammaphi}), in which case we get
\be
\gamma^{\rm (IR)} \simeq \frac{l_{10}  \lambda^2}{480a^3 c^2}\frac{\mu^2}{\Lambda^2}\,, \ \ \ \  \mu\ll {c \over a}\ \Lambda \,.
\label{betaIRLif}
\ee
Equation (\ref{betaIRLif}) implies that the RG evolution of $\gamma$ in the IR is essentially frozen, in complete analogy to what happens in standard Lorentz invariant theories
below the mass scale.
Equation (\ref{betaIRLif}) is best
understood by noticing that eq.(\ref{MOMLif2}) 
is well-behaved in the IR, due to the $c^2 k^2$ term that acts as an IR regulator
and forbids the presence of any IR singularity for $k\rightarrow 0$. Like in the usual Lorentz-invariant $\phi^4$ theory considered before, the IR finiteness is responsible 
for the freezing of the coupling. The essential aspects of the above results  should be  general and apply to any weighted marginal operator in Lifshitz-type theories.

A computation very similar to the one above (but algebraically more involved)  gives
\be
\beta_\lambda^{\rm (IR)}  \simeq 
k_\lambda \frac{\lambda^2}{a^3c^2}
\frac{\mu^2}{\Lambda^2} \lambda\,, \ \ \ \
\beta_{a^2}^{\rm (IR)} \simeq   \frac{2k_a \lambda^2}{a^3c^2}
\frac{\mu^2}{\Lambda^2} a^2 \,, 
 \ \ \ \beta_{c^2}^{\rm (IR)} \simeq \frac{2k_c \lambda^2}{a^3 c^2}
\frac{\mu^2}{\Lambda^2} c^2\,,
\label{betaphi3ac}
\ee
where $k_{\lambda}=-7 l_{10}/480 $, $k_a=183 l_{10}/22400 $ and $k_c= -3l_{10}/400$. We have checked that $\beta_\lambda^{\rm (UV)}$, $\beta_{a^2}^{\rm (UV)}$ and $\beta_{c^2}^{\rm (UV)}$ completely agree with those  found in the MS scheme in section 4, eqs.(\ref{RGphi3}) and (\ref{betacphi3}). Notice that it is crucial to take $p^0\propto \mu^2$ in the UV regime to get agreement
with the MS scheme. By taking $p^0\propto \mu$, we would have obtained a mismatch by a factor 1/2 in the $\beta^{(UV)}$-functions.
The effective IR coupling is 
\be
y  \equiv \frac{\lambda^2}{a^3 c^2} \, ,
\ee
to be compared with the UV effective coupling $x=\lambda^2/a^5$. Contrary to $x$, however, $y$ is not well-defined, in the sense that it sensitively depends on the choice of subtraction point, i.e. the choice of $\mu$. This is obvious, considering that in the IR the $\beta$-functions present an explicit dependence on $\mu$. It is straightforward to solve the RG equations in the IR. Restoring $\Lambda$, we get
 \bea
 y(\mu) \a =\a  \frac{y(\Lambda)}{1+k_yy(\Lambda) (1-\frac{\mu^2}{\Lambda^2})/2}\,, \nn \\
  \lambda(\mu) \a = \a \lambda(\Lambda) \bigg( \frac{y(\mu)}{y(\Lambda)} \bigg)^{\frac{k_\lambda}{k_y}}\,, \nn \\
 a^2(\mu) \a = \a a^2(\Lambda) \bigg( \frac{y(\mu)}{y(\Lambda)} \bigg)^{\frac{2k_a}{k_y}}\,, \nn \\
 c^2(\mu) \a= \a c^2(\Lambda) \bigg( \frac{y(\mu)}{y(\Lambda)} \bigg)^{\frac{2k_c}{k_y}}\,,
\label{RGIRsolved}
  \eea
 where $k_y=2k_\lambda-2k_c-3k_a$. Eqs. (\ref{RGIRsolved}) exhibit the fact that in the IR regime the RG evolution of $a^2$, $c^2$ and $\lambda$, as induced by $\lambda$ itself,  freezes.  More precisely, for $\mu\ll c\Lambda /a$ --- modulo a small threshold effect shifting the IR values of the couplings from the UV ones --- the energy-dependence of the low-energy couplings is proportional to $(\mu/\Lambda)^2$.

\subsection{IR effects of relevant couplings }
 
In the previous subsection we have shown that weighted marginal operators become inoperative for $\mu\ll c \Lambda /a$. 
This is in some sense expected, since in the IR one should recover the usual classification of operators in terms of canonical rather than weighted dimensions. What is weighted marginal in the UV 
becomes  standard irrelevant in the IR. However, care has to be paid to the weighted relevant operators which become standard marginal in the IR. These operators deserve a separate discussion. They are obviously negligible in the UV but  they can efficiently transmit the UV Lorentz violation to the IR theory. Indeed, as we will  see, Lorentz-symmetry-breaking  weighted relevant operators which become standard marginal in the IR will in general mediate Lorentz violation effects to the whole IR Lagrangian. 
 
 In order to illustrate the effect, we consider a simple IR toy model in 3+1 dimensions, consisting of 
a fermion interacting with a scalar by means of a Yukawa interaction.
We imagine that, say, the scalar is described at high energy by a Lifshitz-like dynamics with some $z>1$.
The Yukawa coupling is standard marginal in the usual sense, but weighted relevant.
We want to study its effect in the IR where, as we have just shown, the weighted marginal couplings of the Lifshitz theory have a negligible beta function. We assume that the only remnant of the non-Lorentz invariance of the UV completed theory is a tiny difference in the speed of light of the fermion and scalar:
$c_\psi=1$, $c_\phi = 1 + \delta c $, with $\delta c  \ll 1$. The Lagrangian is simply
\be
{\cal L} = \frac 12 (\dot \phi)^2 - \frac{c_\phi^2}2 (\partial_i \phi)^2 + \bar \psi (\dslash_0 - c_\psi \vec{\dslash} \psi) - g \bar \psi \psi \phi \,.
\ee
We will now show how, due to a non-vanishing value of $\delta c $, a logarithmic running in $c_\psi$ is induced.
We work at linear order in $\delta c $ and in the MS scheme. We omit details, since the one-loop graphs of this model are straightforward and can be found in standard textbooks.
We denote by $Z_\phi$ and $Z_\psi$ the wave-function renormalization constants of $\phi$ and $\psi$, 
associated with the renormalization of $(\dot \phi)^2$ and  $\bar \psi\, \dslash_0 \psi$, by 
$Z_{c\phi}$ and $Z_{c\psi}$  the wave-function renormalization constants  
associated with  $(\partial_i \phi)^2$ and $ \bar \psi\, \vec{\dslash} \psi$ and by $Z_g$ the vertex renormalization constant. 
We get 
\bea
Z_{c\psi} Z_\psi^{-1}  \a =  \a 1-{g^2\ov (4\pi)^2}{1\ov \epsilon}\Big(\frac 23 \delta c +{\cal O}(\delta c ^2)\Big) \nn \\
Z_{c\phi} Z_\phi^{-1} \a = \a 1-{g^2\ov (4\pi)^2}{1\ov\epsilon}\Big(8 \delta c +{\cal O}(\delta c ^2) \Big)\,, \nn \\
Z_g \a = \a 1+{g^2\ov (4\pi)^2}{1\ov\epsilon}\Big(2+{\cal O}(\delta c) \Big) \,.
\label{allZ}
\eea
{}From eqs.(\ref{allZ}) we get
\be 
\beta_{c_\psi}  =  -\frac{g^2}{24\pi^2} \delta c  \,, \ \ \ \ 
\beta_{\delta c}   =  \frac{5g^2}{24\pi^2} \delta c  \,, \ \ \ \ \ 
\beta_g  =  \frac{5g^3}{16\pi^2} \,.
\ee
Defining $\alpha \equiv g^2/(4\pi)$, we finally get the following RG evolution for $c_\psi(t)$, $\delta c (t)$ and $\alpha(t)$:
\bea
\alpha(t) \a  = \a \frac{\alpha_0}{1-\frac{5}{2\pi} \alpha_0 t} \,, \nn \\
\delta c (t) \a = \a \delta c _0 \bigg(\frac{\alpha(t)}{\alpha_0} \bigg)^{\frac{7}{15}} \,, \nn  \\
c_\psi(t) \a = \a c_{\psi,0} -\frac{\delta c _0}{7} \bigg[\bigg(\frac{\alpha(t)}{\alpha_0} \bigg)^{\frac{7}{15}}-1 \bigg]\,.
\label{RGIRYukawa}
\eea
Equations (\ref{RGIRYukawa}) shows that any small Lorentz symmetry breaking term in the IR theory (coming from the underlying UV theory) induces, by quantum effects,
an energy-dependent speed of light for all particles sensible to the breaking term. From this example it should also be clear that the effect is general.

\section{Conclusions}

In this work we have carried out  the analysis of the one-loop RG evolution of Lifshitz-like theories, by mainly focusing on  two specific Lorentz violating scalar field theories.
Our primary interest was to look for possible simple mechanisms (see also \cite{nielsen}) which would alleviate the fine-tuning otherwise needed in these theories to recover Lorentz symmetry at low energies with the needed accuracy.
We have focused our attention on two particular  operators, the spatial kinetic terms $c_{1,2}^2(\vec\partial \phi_{1,2})^2$.
There are essentially two regimes of interest. In the UV Lifshitz-regime, $c^2$ and $\delta c^2$ logarithmically run with the energy,
and the running is governed by weighted marginal operators. In the IR, if all sources of Lorentz symmetry breaking are vanishing, one recovers
Lorentz invariance, which clearly forbids any scale-dependence for $c^2$. However, due to the effects of standard marginal couplings, any small Lorentz symmetry breaking term in the theory leads to phenomenologically unacceptable logarithmic dependence of $c^2$ on the energy scale.  As it has already been pointed out, 
our considerations seem to be very general and independent of additional structure and/or symmetry that would be present in more realistic theories, such as gauge symmetries. It should be clear that, although we have been focusing on the
spatial kinetic term operator,  our considerations may be equally applied to other  Lorentz violating parameters associated with other operators in a theory (this could lead to the need of additional fine-tuning).

In section 5.3 we computed the evolution of $c$ in the IR in a simple low-energy scalar-fermion theory, assuming the presence of a small 
$\delta c$, coming from the non-Lorentz invariant UV theory. In this example $\delta c$ is driven to small values at low energies, see eq. (\ref{RGIRYukawa}).
A logical possibility is that in the Standard Model the RG flow could similarly drive the $\delta c_{ij}\equiv c_i-c_j$ corresponding to
 any  $i,j$ pair of particles to sufficiently small values,  below the existing phenomenological bounds.
Although not completely excluded by our analysis (since we are not computing RG in a Lifshitz-type Standard Model), this possibility seems unlikely, since the logarithmic running
of $\delta c$ at low energies is too slow to drive a $\delta c$ of order 1 at the scale $\Lambda$ to values  compatible with the stringent existing bounds $\delta c< 10^{-(21\div 23)}$. 
Moreover, in a UV free theory, the same mechanism of section 5.3 would give a $\delta c$ that increases towards the IR.

An order of magnitude of other potential phenomenological problems can also be obtained by applying our considerations
to the photon in a weighted renormalizable version of QED. 
At low energies we can neglect higher orders terms in $p$ (always present in Lifshitz type theories),  so that
the photon dispersion relation will be typically of the form
\be
\omega^2 = c^2_\gamma(t) p^2 \,,
\ee
where $\omega$ is the energy and
\be
c_\gamma^2(t) = c^2_0 \big(1- f t\big)^{r}\,,
\label{Effc2}
\ee 
with some unspecified constants $f$ and $r$. 
The behavior
eq.(\ref{Effc2}) should be induced by one loop effects as in section 5.3 once a small $\delta c$ between the photon and a charged particle is introduced in the Largrangian.
This behavior can be compared with known experimental constraints.
An interesting constraint on the energy dependence of $c_\gamma^2$ 
at moderately high energies has been recently given by the FERMI experiment, detecting the photon spectrum of the gamma ray burst
GRB 080916C at red-shift $z=4.35$ \cite{Abdo:2009zz}\footnote{We are grateful to David Mattingly for bringing to our attention \cite{Abdo:2009zz} and providing us the estimate that follows.}. By measuring the time delay ($\sim 10$ sec) between the ``low-energy" ($\lesssim 1$ MeV) component  of the burst with respect to the high energy one ($\sim $ 10 GeV),  we get the following rough 
experimental bounds\footnote{In \cite{Chen:2009ka}, the same time delay observed by FERMI in the photon spectrum emitted by GRB 080916C was ``explained" in the context of the Ho$\breve{{\rm r}}$ava theory of gravity, by means of the higher derivative terms appearing in the photon dispersion relation. They did not consider the one-loop effect studied here.}
\be 
|c^2({\rm 1MeV}) - c^2(10{\rm GeV})| \lesssim 10^{-17}\,.
\label{cexp}
\ee
By taking $r\sim 1$ in eq.(\ref{Effc2}), eq.(\ref{cexp}) approximately gives the constraint on $f$ described in eq.(\ref{cgammaConstr}).
Similar (although milder) bounds  exist for other particles as well.

One might think that, in the context of the Ho$\breve{{\rm r}}$ava  theory of gravity, the bounds we are finding can be evaded for 
all the Standard Model (SM) particles by simply not introducing the higher derivative spatial interactions responsible for the effects
studied in this paper. Being the SM renormalizable, there is no real need of introducing them.
Aside from the lack of a clear principle besides this choice, radiative effects induced by graviton loops should nevertheless generate these Lorentz-violating couplings in the SM sector at some loop order.

Finally, let us conclude by noticing that the one-loop corrections to $c^2$ we considered are quadratically divergent. If new physics above the scale $\Lambda$ is assumed, this can give rise to a  naturalness problem, similar to the standard one affecting the Higgs boson mass.

\section*{Acknowledgments}

We would like to thank Stefano Liberati and especially David Mattingly for useful discussions on phenomenological bounds on Lorentz violating theories.  We also thank E. Kiritsis for useful comments. M.S. thanks the Galileo Galilei Institute (GGI) in Florence for hospitality, where part of this work has been done, and INFN for
partial support.

\end{document}